# CLOUD PENETRATION TESTING

Ralph LaBarge[1] and Thomas McGuire[2]

[1]Johns Hopkins University Applied Physics Laboratory, Laurel, MD, USA
`Ralph.LaBarge@jhuapl.edu`
[2]Johns Hopkins University, Baltimore, MD USA
`Tmcguir3@jhu.edu`

## ABSTRACT

*This paper presents the results of a series of penetration tests performed on the OpenStack Essex Cloud Management Software. Several different types of penetration tests were performed including network protocol and command line fuzzing, session hijacking and credential theft. Using these techniques exploitable vulnerabilities were discovered that could enable an attacker to gain access to restricted information contained on the OpenStack server, or to gain full administrative privileges on the server. Key recommendations to address these vulnerabilities are to use a secure protocol, such as HTTPS, for communications between a cloud user and the OpenStack Horizon Dashboard, to encrypt all files that store user or administrative login credentials, and to correct a software bug found in the OpenStack Cinder type-delete command.*

## 1. INTRODUCTION

This paper discusses penetration testing of the OpenStack Essex Cloud Management Software package. The paper is organized into nine sections including (I) Introduction, (II) OpenStack Cloud Management Software, (III) Selection of Penetration Testing Software, (IV) Design & Implementation of the Test Cloud, (V) Design & Implementation of the Penetration Test Environment, (VI) Description of the Penetration Tests Performed (VII) Test Results, (VII) Summary and Conclusions, and (IX) References.

## 2. OPENSTACK CLOUD MANAGEMENT SOFTWARE

OpenStack includes four core services, and a set of ancillary services, which provide an integrated cloud management environment. Core services include "Compute", "Storage", "Networking" and "Dashboard". Shared services include "Identity" and "Image". Figure 1 shows a block diagram of the OpenStack cloud management software.

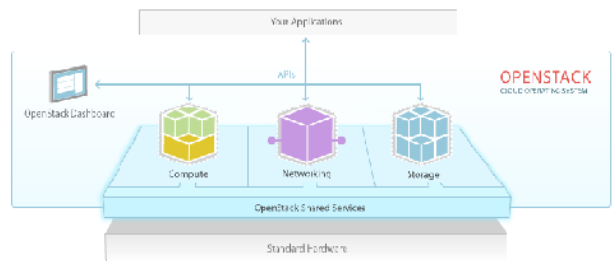

Figure 1. OpenStack Cloud Management Software

## 2.1. OpenStack Compute (Nova)

OpenStack Compute is used to provision and manage large networks of virtual machines. Common use cases for OpenStack Compute include public cloud service providers offering Infrastructure as a Service (IaaS) cloud services, IT departments offering private cloud services within their organizations, Big Data applications using tools like Hadoop, and High-performance computing (HPC) applications. A partial list of OpenStack Compute features includes:

- Manage virtualized commodity server resources including CPU, memory, disk, and network interfaces
- Manage local area networks including Flat, Flat DHCP, VLAN DHCP, IPv4 and IPv6 networks
- Virtual Machine image management services to store, import, share, and query virtual images
- Ability to assign (and re-assign) floating IP addresses to VMs
- Role Based Access Control (RBAC) provides security by user, role and project
- VM Image Caching on compute nodes provides faster provisioning of VMs

## 2.2. OpenStack Storage (Swift & Cinder)

OpenStack Storage provides both object and block storage for use with servers and applications. Object storage is a distributed storage system for static data such as virtual machine images, backups and archives. Objects and files are written to multiple disk drives spread throughout the OpenStack cloud, providing scalability and redundancy. OpenStack also provides persistent block level storage devices for use with compute instances that require high performance storage for databases, expandable file systems, or a server that requires access to raw block level storage.

A partial list of OpenStack Storage features includes:

- Use of commodity hard drives to reduce the cost per storage byte
- Self-healing: Data is copied to several different places across the cloud making the storage system highly redundant and reliable
- Unlimited storage with both horizontal and vertical scaling
- Very large scale: multiple Petabytes with billions of individual objects
- Amazon S3 (Elastic Block Storage) API support
- Built in management utilities provide account management, container management and storage monitoring functions

### 2.3. OpenStack Networking (Quantum)

OpenStack Networking is an API-driven system for managing cloud networks and IP addresses. A partial list of OpenStack Networking features includes:

- Manages IP addresses, allowing for static, DHCP or floating IP addresses
- Several networking models including flat networks or VLANs
- Allows users to create and manage their own networks
- Support for software-defined networking technology (i.e. OpenFlow)
- Network framework allows for a variety of devices to be integrated into the cloud including intrusion detection systems, load balancers, firewalls, etc.

### 2.4. OpenStack Dashboard (Horizon)

OpenStack Dashboard allows cloud administrators and users to provision, manage and control cloud compute, storage and networking resources. Cloud administrators use the dashboard to create users and projects, assign users to projects, and set limits on the resources for those projects. Cloud users can also use the dashboard to provision and control the resources that have been allocated to their projects. The OpenStack Dashboard is implemented as an extensible web-based application.

### 2.5. OpenStack Identity (Keystone)

OpenStack Identity maintains a database of users and maps these users to the OpenStack services they are allowed to access. It provides a common authentication system across the cloud and can be integrated with third party backend directory services (i.e. LDAP). Multiple forms of authentication are supported including standard username and password credentials, token-based systems and Amazon Web Services style logins. OpenStack Identity allows cloud administrators to set common policies across users and systems, to create users and tenants, and to define permissions for compute, storage and networking resources.

### 2.6. OpenStack Image (Glance)

The OpenStack Image Service provides discovery, registration and delivery services for disk and server images. Cloud administrators can create base image templates from which cloud users can create new instances. Users and administrators can also create and store snapshots of images. Images can be stored in a variety of common formats including Raw, VHD (Hyper-V), VDI (VirtualBox), qcow2 (Qemu/KVM), VMDK (VMware), and OVF (VMware, others).

## 3. SELECTION OF PENETRATION TESTING SOFTWARE

Penetration testing software is used to evaluate the security of a computer system or network by simulating an attack. The simulated attack can come from an outsider (e.g. a hacker) or an insider (e.g. a disgruntled employee). Several penetration testing techniques will be used in this research effort, including fuzzing, session hijacking, and credential theft.

### 3.1. Fuzzing

Fuzzing is used in computer security to describe a number of tools and techniques used to discover vulnerabilities by subjecting a program to a wide variety of inputs. Computer programmers, and testers have used fuzzing techniques since the early 1970's. [1] The term

"fuzzer" was first used in 1988 by Barton Miller, a professor at the University of Wisconsin-Madison (UW-M). Miller, his associates, and students from his Computer Science classes at UW-M developed a series of fuzzers to test the reliability of UNIX system routines and application programs. [2]

Another milestone in the history of fuzzing was the initial release of SPIKE in 2001, and its subsequent presentation at Black Hat 2002 by Dave Aitel of Immunity, Inc. [3] SPIKE is a fuzzing framework that allows a tester to define the structure of a program's input as a series of layered blocks. Understanding the structure of the input stream allows fuzzing to be more efficient than simply generating random input data and providing it to a program under test. For example if a program's input includes a checksum, generating completely random input data to fuzz the program would be extremely inefficient since the random input data would likely not include a valid checksum, and would thus be rejected by the program.

Grammar based fuzzing is a combination of random fuzzing techniques with block-based fuzzing techniques. A minimal definition of the protocol to be fuzzed is created to automatically generate inputs to the program under test that partially complies with the protocol specification. Critical protocol parameters, such as checksums, can be completely specified, while less important parameters can be randomized. An example of a grammar-based fuzzer is the PROTOS project developed at the University of Oulu in Finland. [4]

Since 2002 the popularity of fuzzing has grown, as has the sophistication and number of open-source and commercial fuzzing tools. Today, fuzzing is widely recognized as a valid computer security test method, and is being used by many commercial software development companies. Microsoft uses "white-box" fuzzing as part of their quality assurance process. Dr. Patrice Godefroid of Microsoft defines white-box fuzzing as "a new approach to fuzzing pioneered at Microsoft in the SAGE tool and based on symbolic execution and constraint solving techniques." [5] According to Godefroid a Windows 7 test team found 50% more bugs using a white-box fuzzer (SAGE) than all other traditional fuzzers combined.

### 3.1.1. Comparison of Fuzzing Tools

While there are dozens of fuzzing tools available, the authors have chosen to select between a subset of open-source tools for this project. Open-source tools that will be considered include BED, SFUZZ, SICKFUZZ, and SPIKE.

BED, also known as Bruteforce Exploit Detector, is a protocol fuzzer developed by Martin Muench & Eric Sesterhenn that detects common vulnerabilities including buffer overflows, format string bugs, and integer overflows. [6] BED supports fuzzing the FINGER, FTP, HTTP, IMAP, IRC, LPD, PJL, POP, SMTP, SOCKS4 and SOCKS5 protocols. BED is an open-source Linux based fuzzing tool that is relatively easy to use. BED does not offer any options to perform command line fuzzing.

SFUZZ, also known as Simple Fuzzer, is a block-based fuzzer that includes a number of predefined scripts for popular protocols including HTTP, POP3, RTSP, SMTP and Twitter. [7] SFUZZ can also be used to perform command line fuzzing. SFUZZ is an open-source Linux based fuzzing tool that is very easy to use for the predefined protocol scripts, and moderately easy to use for fuzzing command lines.

SICKFUZZ is a Python front-end for the SPIKE fuzzing tool. SICKFUZZ is designed to fuzz the HTTP protocol, and includes six predefined test cases for HTTP functions (HEAD, GET, POST, etc.). [8] SICKFUZZ is an open-source Linux based fuzzing tool that is relatively easy to use, although it generated a large number of run-time errors when the authors attempted to use it to fuzz the OpenStack Horizon HTTP interface.

SPIKE is a block-based fuzzer developed by David Aitel of Immunity, Inc. [9] SPIKE is an open-source Linux based fuzzing tool that is relatively hard to use since it requires a strong knowledge of the C programming language, as well as detailed knowledge of the protocols that are fuzzed. SPIKE is probably the most powerful open-source block-based fuzzer available today.

Based on analysis of how each fuzzing product met the specific requirements for this research effort, as well as hands-on testing or live product demonstrations of each product, the authors chose to use BED and SFUZZ for the test phase of this research effort. BED will be used to fuzz the OpenStack Horizon user interface via a network protocol (HTTP) interface. SFUZZ will be used to fuzz the OpenStack Horizon user interface via a network protocol (HTTP) interface. SFUZZ will also be used to fuzz the OpenStack command line interfaces.

### 3.2. Session Hijacking

Session hijacking involves the exploitation of a valid session key to gain unauthorized access to a computer system or a computer network. There are four basic types of session hijacking attacks including session fixation, session sidejacking, session key theft, and cross-site scripting. The session sidejacking technique will be used in this research effort. Session sidejacking uses packet sniffing tools to capture a login sequence, and thus gain access to the user's session key. Figure 2 shows a block diagram of a session hijacking attack.

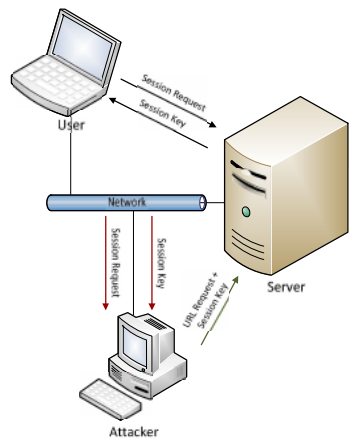

Figure 2. Session Hijacking

As shown in Figure 2, the user initiates an HTTP session with a server over a network connection. The user's web browser sends a session request to the server, which responds with a session key. The session key will be used in subsequent communication as an alternative to the user's login credentials. The attacker gains access to the session key by packet sniffing on the network the user and server use to communicate. Once the session key has been sniffed, the attacker can use a web browser to access restricted web pages on the server using the stolen session key.

Penetration testing tools developed by Robert David Graham of Errata Security called Ferret and Hamster will be used to perform the session sidejacking tests on the OpenStack cloud. [10] Ferret sniffs and captures session keys transmitted over a network connection. Ferret stores the URLs of web pages along with the session keys associated with those pages in a text file. Hamster reads data from a Ferret text file, opens a standard web browser, and provides an easy to use graphical interface where an attacker can click on any of the captured URLs and navigate the browser directly to restricted web pages using the captured session keys.

### 3.3. Credential Theft

Credential theft is a relatively simple penetration technique where an attacker steals, or guesses, a user's login credentials. User credentials can be stored in unencrypted files on the computer's hard drive, or transmitted over an unencrypted network connection. In either case, once an attacker has gained access to the unencrypted user credentials they can use them to impersonate the user and gain access to their restricted data. Another technique used to steal user credentials is through the use of a key logging program that is used to remotely login to a computer. The most common security incident at the National Energy Research Scientific Computing Center (NERSC) is account compromises resulting from credential theft. [11]

For this research effort Wireshark will be used to monitor the network connection between an OpenStack user and the OpenStack server. [12] Captured packets will be analyzed to determine if user credentials have been transmitted over the network as unencrypted data.

Analyzing the OpenStack server's hard drive to locate unencrypted user login credentials will be performed using a number of standard Linux file access and editing tools. For example gedit can be used to open and search through text based configuration files to look for user names, passwords or other forms of login credentials.

## 4. DESIGN & IMPLEMENTATION OF THE TEST CLOUD

This section will describe how the OpenStack cloud server was built and configured prior to the start of the testing effort.

OpenStack (Essex) cloud management software was installed on an Ubuntu 12.04 LTS system with dual quad-core Intel i7-3770 processors operating at a clock speed of 3.4 GHz. The OpenStack server included 16 GB of system RAM, a 3 TB local hard drive, and two Gbit Ethernet network interfaces. One of the network interfaces was used to connect the OpenStack server to an Internet Gateway, while the other was used to provide network connectivity between the OpenStack server and the various computers used to perform vulnerability tests.

Figure 3 shows a high-level block diagram of the OpenStack Test Cloud as well as the various computers that were used to perform the penetration tests

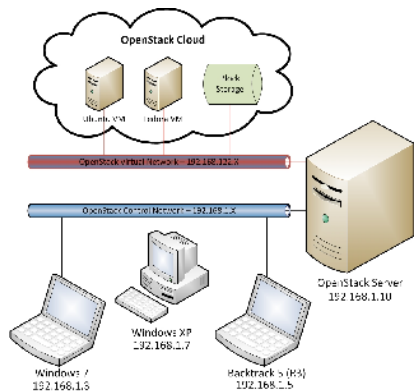

Figure 3. OpenStack Test Cloud Implementation

The OpenStack software was downloaded from the DEVSTACK web site, which offers shell scripts to automatically install OpenStack on a variety of target platforms. [13] The DEVSTACK "All-in-one" shell script was used to install OpenStack on the target server. Using the shell script from DEVSTACK reduced the OpenStack installation process to the following three steps:

1. Install Ubuntu 12.04 LTS on the target server
2. Clone the DEVSTACK installation software from Github
3. Deploy the OpenStack cloud software by executing the DEVSTACK shell script

The OpenStack installation process using the DEVSTACK shell script took approximately two hours. A few additional configuration steps were required after the installation process had been completed, such as editing configuration files to include the correct network addresses. Additional configuration steps included downloading virtual machine images for use in the OpenStack cloud, creating test user accounts, creating virtual machine instances, creating volumes, and associating volumes with specific virtual machine images. Most of the cloud configuration steps can be performed from a remote workstation using the OpenStack Horizon Dashboard, or they can be performed from the command line on the OpenStack server.

While the primary purpose of this research effort was to detect vulnerabilities in the OpenStack cloud management software, it was desirable to configure the OpenStack test system so that it was fully operational, with multiple user accounts, multiple virtual machine images, and multiple volumes. CentOS, Fedora, RedHat, ttylinux and Ubuntu virtual images in the QCOW2 format were downloaded and installed on the OpenStack server. These OpenStack compliant virtual machine images are available for download from Rackspace Cloud Builders. [14]

## 5. DESIGN AND IMPLEMENTATION OF THE PENETRATION TEST ENVIRONMENT

This section will describe how the systems used to perform OpenStack penetration tests were built and configured prior to the start of the testing effort.

As shown in Figure 3, three different computers were configured to support penetration testing during the research effort. A Windows 7 laptop was used primarily to connect remotely to the OpenStack Horizon Dashboard interface. A Windows XP system was configured with a promiscuous mode network interface card, Wireshark, and a few other tools to analyze network traffic to and from the OpenStack server.

A Backtrack 5 (R3) system was configured with a variety of penetration tools, including the fuzzing tools discussed earlier. Some penetration tools were also installed on the OpenStack server itself in order to facilitate command line fuzzing. Backtrack 5 (R3) is a Linux based penetration testing tool that is used by cyber security professionals. Backtrack 5 (R3) includes hundreds of different cyber security analysis and penetration testing tools and is available as a free download. [15] All the open-source fuzzers discussed earlier in this paper are available in the Backtrack 5 (R3) release. A number of network scanners, including Zenmap, and network packet capture tools, including Wireshark, are also available in the Backtrack 5 (R3) release.

Prior to performing penetration tests, a detailed network scan of the OpenStack server was performed using the Zenmap program on the Backtrack 5 (R3) system. Zenmap is a graphical user interface for the nmap program. [16] Nmap, also known as Network Mapper, is an open source utility for network discovery and security auditing. The result of the scan indicates that there are 19 network ports on the OpenStack server that could be used as attack vectors. Table 1 lists the open network ports that are used by OpenStack. [17]

Table 1. OpenStack Network Ports

| Port 80 – HTTP | Port 5672 – AMQP | Port 8776 – Nova API |
| --- | --- | --- |
| Port 3260 – Glance API | Port 5800 – X11VNC | Port 9191 – Glance API |
| Port 3306 – MySQL | Port 5900 – VNC | Port 9292 – Glance API |
| Port 3333 – Nova API | Port 8773 – EC2 API | Port 35357 – Keystone API |
| Port 4369 - EPMD | Port 8774 – EC2 API | |
| Port 5000 – Keystone API | Port 8775 – Nova API | |

Once the network ports used by OpenStack were identified, a series of tests were run to characterize the network packet structures used for each port. These were fairly simple tests that involved monitoring the network using Wireshark while using the Horizon Dashboard to configure different aspects of OpenStack.

## 6. DESCRIPTION OF THE PENETRATION TESTS PERFORMED

This section will describe the specific penetration tests that will be performed on the OpenStack cloud management software.

### 6.1. OpenStack Horizon HTTP Fuzzing

The OpenStack Horizon Dashboard provides administrators and users a graphical interface to access, provision and automate cloud-based resources. The Horizon Dashboard services use a standard Apache web server listening on port 80. This portion of the OpenStack penetration testing effort will attempt to detect vulnerabilities in the OpenStack Horizon Dashboard and its associated Apache web server using the Bruteforce Exploit Detection (BED) and Simple Fuzzer (SFUZZ) penetration test tools.

**6.1.1 Bruteforce Exploit Detection (BED) HTTP Fuzzing**

The Backtrack 5 R3 system will be used to run the *BED* program to test the OpenStack Horizon Dashboard's HTTP service.

For HTTP testing of the OpenStack Horizon Dashboard service *BED* will be invoked from the Backtrack 5 R3 command line as follows:

*perl bed.pl –s HTTP –t 192.168.1.10 –p 80 –o 2*

This command instructs the *BED* program to fuzz the HTTP service listening on 192.168.1.10:80 with approximately two seconds between each fuzzing attempt. The *BED* program will transmit a series of fuzzed packets to test the HTTP HEAD, GET, POST, User-Agent, Host, Accept, Connection, Referer, Authorization, From, Charge-to, If-Modified-Since, and Pragma functions.

**6.1.2 Simple Fuzzer HTTP Fuzzing**

The Backtrack 5 R3 system will also be used to run the *sfuzz* program to test the OpenStack Horizon Dashboard's HTTP service. *sfuzz* includes a basic HTTP fuzzing configuration file. Configuration files are used to specify how the *sfuzz* program generates the fuzzed packets that are sent to the HTTP service.

International Journal on Cloud Computing: Services and Architecture (IJCCSA),Vol.2, No.6, December 2012For penetration testing of the OpenStack Horizon Dashboard HTTP service *sfuzz* will be invoked from the Backtrack 5 R3 command line as follows:

> ./sfuzz –T –f scripts/basic.http –S 192.168.1.10 –p 80

This command instructs the SFUZZ program to fuzz the HTTP service listening on 192.168.1.10:80 using the "basic.http" fuzzing configuration file located in the scripts directory. This configuration file instructs the SFUZZ program to generate a series of fuzzed HTTP GET, HEAD and POST commands with a fuzzed string length between 1 and 10,024 characters.

## 6.2. OpenStack Command Line Fuzzing

OpenStack includes six services that have command line functions with at least one parameter. These services include Cinder, Glance, Keystone, Nova, Quantum and Swift. Each command with at least one required parameter from these six services will be fuzzed using the SFUZZ program. OpenStack has been designed to require user authentication for most commands. During the SFUZZ tests for each command that requires user authentication the fuzzed command line will include a valid OpenStack user name, password, tenant name, and a URL pointing to the Keystone server that will validate these credentials.

The OpenStack credentials used for the command line fuzzing are provided below. This information is available from the OpenStack Horizon Dashboard by logging into an account, clicking on the "settings" button located in the top right portion of the screen, clicking on the "OpenStack API" button on the left portion of the next screen, and then clicking on the "Download RC File" button at the bottom of the next screen. The OpenStack RC file for the admin account includes the following user credential information required to execute commands.

> OS_TENANT_ID=84a0eb4cabc441ab9325c42f0c6f57a5
> OS_TENANT_NAME="admin"
> OS_USERNAME=admin
> OS_AUTH_URL=http://192.168.1.10:5000/v2.0

The password for admin is not included in the OpenStack RC file, and is required by Keystone for authentication before a command is executed. The password for admin is "adminpassword", which will be included as a command line parameter. The admin account credentials will be used for command line fuzzing since a user with administrative privileges is required to execute some OpenStack commands.

For each of the six OpenStack services a custom SFUZZ configuration file was developed which includes test cases for each OpenStack command with proper user authentication parameters, and one or more fuzzed parameters.

The goal of fuzzing each of these OpenStack commands is to determine if each parameter is properly validated before the command is executed. These fuzzing tests may discover vulnerabilities if the code associated with a command does not properly validate a parameter that has been accepted from the command line.

For each OpenStack service a custom SFUZZ configuration script was created with individual test cases for each unique command and sub-command combination. These configuration files are saved as .txt files. An exemplar test case from the nova.txt file is shown below.

> # Nova Test Case 03
> # Fuzz nova add-fixed-ip (two arguments)

51

```
nova --os_username admin
    --os_password adminpassword
    --os_tenant_name "admin"
    --os_auth_url http://192.168.1.10:5000/v2.0
    add-fixed-ip FUZZ FUZZ
--
```

This test case generates a series of nova *add-fixed-ip* commands which each requires two parameters. Each parameter will be fuzzed to create a series of valid nova *add-fixed-ip* commands with what are most likely invalid parameters.

Each configuration file is processed by the SFUZZ program to create a series of fuzzed commands that are stored in a shell script file. An exemplar shell script line from the nova.sh file is shown below.

```
# Nova Test Case 03
# Fuzz nova add-fixed-ip (two arguments)
nova --os_username admin
    --os_password adminpassword
    --os_tenant_name "admin"
    --os_auth_url http://192.168.1.10:5000/v2.0
    add-fixed-ip AAAA AAAA
```

This shell script command includes two parameters that the SFUZZ program has filled in with a series of four "A" characters. The length of the fuzzed parameter strings generated by the SFUZZ program is controlled by the configuration file. For the OpenStack Cinder, Glance, Keystone, Nova, Quantum and Swift testing efforts the maximum string length of a fuzzed parameter was set to 1025 characters. Table 2 shows the total number of unique fuzzed shell script commands created to test each of the OpenStack services.

Table 2: OpenStack Command Line Fuzzing Tests

| Service Name | Test Cases | Total # of Tests |
|---|---|---|
| Cinder | 16 | 229,600 |
| Glance | 11 | 157,850 |
| Keystone | 24 | 344,400 |
| Nova | 73 | 1,047,550 |
| Quantum | 28 | 401,800 |
| Swift | 8 | 114,800 |
| **Totals** | **160** | **2,296,000** |

Each SFUZZ configuration file also specifies the characters used to build the fuzzed parameters. These include the following ASCII characters: !, /, 0, 9, :, @, A, Z, [, ', a, z, {, and ~. Each of these characters are used to form test strings between 1 and 1025 characters in length, which are then inserted as parameters of the fuzzed OpenStack commands.

Some OpenStack services include a rate-limiting option that limits the number of commands that can be executed over a specific period of time. During the fuzzed command line testing of the nova service the rate-limit feature was disabled to ensure that all of the commands were processed by the nova service. Disabling the nova rate-limit feature is done by editing the etc/nova/api-paste.ini file. [18]

International Journal on Cloud Computing: Services and Architecture (IJCCSA),Vol.2, No.6, December 2012It should be noted that while the number of unique commands that will be fuzzed during these tests is very large, it is possible to perform a much more comprehensive fuzzing of the OpenStack command lines. Only required parameters are being fuzzed during this research effort. In addition to required parameters most OpenStack commands can also contain optional parameters. Also the command line fuzzing performed during this research effort generated fuzzed parameters using only 10 different ASCII characters, and limited the parameter length to 1025 characters. A more comprehensive OpenStack command line fuzzing test could be designed to use all the required and optional parameters, use all 256 valid ASCII characters, and create longer fuzzed parameters. Going to this extreme would likely increase the total number of fuzzing tests to well over 50,000,000.

**6.3. Session Hijacking**

During this portion of the OpenStack penetration test effort an attempt will be made to hijack an HTTP session using a stolen session cookie. The Ferret program will be used to monitor the network connection between an OpenStack user and the OpenStack server. When the OpenStack user connects to the OpenStack server via the Horizon Dashboard, Ferret will capture the session cookie provided by the server to the user's browser. Ferret stores the stolen session cookie in a text file, along with URL data associated with web pages that were visited by the user during their session.

The Hamster program retrieves the stolen session cookie and URL information from the text file created by Ferret, and allows an unauthorized user to hijack the OpenStack user's HTTP session, and gain access to restricted Horizon Dashboard web pages.

The Ferret and Hamster programs were run on a Windows XP system that has a promiscuous mode network interface card, as shown in Figure 3. A Windows 7 laptop was used to login to the OpenStack server using the Test_User_1 account.

There is a known session hijacking vulnerability in OpenStack Essex and Folsom as described in the National Vulnerability Database CVE-2012-2144. [19] The overview for CVE-2012-2144 states "Session fixation vulnerability in OpenStack Dashboard (Horizon) folsom-1 and 2012.1 allows remote attackers to hijack web sessions via the sessionid cookie". A patch to fix this vulnerability was released by OpenStack in May 2012. [20] The patch rotates session cookies after a user logs out, and properly clears sessions. [21]

This patch revised the following six python scripts used in portions of the Horizon login process:

- horizon/exceptions.py,
- horizon/middleware.py,
- horizon/tests/auth_tests.py,
- horizon/users.py,
- horizon/views/auth.py,
- horizon/views/auth_forms.py

The penetration tests will attempt to verify that session cookies are reset after an OpenStack user logs out, and will also attempt to hijack a user's session before they logout.

**6.4 Credential Theft**

During this portion of the OpenStack penetration test effort an attempt will be made to steal user credentials transmitted over the network connection, or stored in files on the OpenStack server.

53

International Journal on Cloud Computing: Services and Architecture (IJCCSA),Vol.2, No.6, December 2012

The Wireshark program will be used to monitor the network connection between an OpenStack user and the OpenStack server. When the OpenStack user connects to the OpenStack server via the Horizon Dashboard, Wireshark will attempt to capture the user's login credentials. During this penetration test a Windows 7 computer will be used to login to the OpenStack Horizon Dashboard using Test_User_1's login credentials.

In addition to attempting to steal user credentials transmitted over the network connection, the OpenStack server will be analyzed to determine if any user login credentials are stored in files on the server. No special penetration test programs are required for this analysis. Standard Linux programs like gedit can be used to analyze OpenStack files to determine if user credentials may be stored in them. The OpenStack server directories that will be analyzed include the "devstack", "etc", and "var" directories. Each of these directories includes a number of OpenStack configuration files, program files and script files.

## 7. TEST RESULTS

This section will present the results of the penetration tests performed on the OpenStack cloud management software.

### 7.1 OpenStack Horizon Dashboard HTTP Fuzzing

The OpenStack Horizon Dashboard HTTP server was fuzzed using both the BED and SFUZZ penetration test tools.

### 7.1.1 BED HTTP Fuzzing Test Results

Figure 4 is a Wireshark HTTP/Packet Counter Summary from the BED fuzzing test. During this test a total of 93,896 HTTP packets were transmitted, and the test took approximately four hours to complete. Throughout the BED fuzzing test the OpenStack HTTP server was responsive, and did not exhibit any behaviour that would indicate a potential vulnerability.

Figure 4. BED HTTP/Packet Counter          Figure 5: sfuzz HTTP/Packet Counter

International Journal on Cloud Computing: Services and Architecture (IJCCSA),Vol.2, No.6, December 2012

### 7.1.2 SFUZZ HTTP Fuzzing Test Results

Figure 5 is a Wireshark HTTP/Packet Counter Summary from the SFUZZ fuzzing test. During this test a total of 3,474 HTTP packets were transmitted, and the test took approximately ten minutes to complete. Throughout the SFUZZ fuzzing test the OpenStack HTTP server was responsive, and did not exhibit any behaviour that would indicate a potential vulnerability.

### 7.2 OpenStack Command Line Fuzzing

The OpenStack server was command line fuzzed using six different shell scripts created by the SFUZZ penetration test program. These shell scripts tested the OpenStack cinder, glance, keystone, nova, quantum and swift command line APIs.

### 7.2.1 Cinder Command Line Fuzzing Test Results

Figure 6 shows a screen grab taken while the cinder command line fuzzing test was running. The left side of the screen shows a Linux terminal executing the cinder fuzz test shell script, while the right side of the screen shows the Ubuntu System Monitor program, which was used to monitor the CPU, Memory and Network utilization rates during the test. As shown in Figure 6, the average CPU utilization rate was less than 50% during this test.

The OpenStack cinder service did not crash, or exhibit any unexpected behaviors during the command line fuzz testing. However, one problem was observed after the end of the fuzz test. During the fuzz testing over 9000 cinder volume types were created. After the fuzz testing, an attempt was made to delete all of these cinder volume types so that the OpenStack server was in a known state before running the glance fuzz tests.

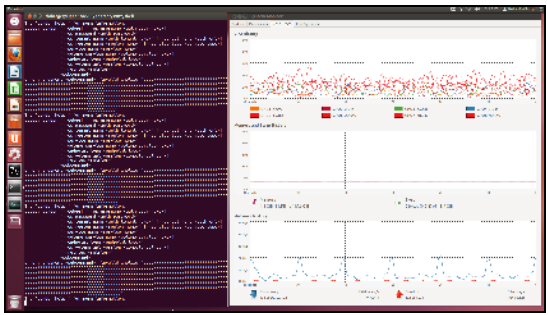
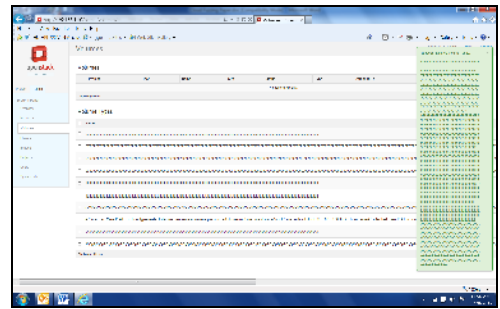

Figure 6: Cinder Fuzz Test          Figure 7: Cinder Failure

The OpenStack Horizon Dashboard was used to delete the volume types created during the cinder fuzz testing. Ten of the volume types could not be deleted. Figure 7 shows a screen grab of the OpenStack Horizon Dashboard showing the ten volume types that could not be deleted after the cinder fuzz tests.

The green box on the right side of Figure 7 is a dialog box confirming that all ten volume types were successfully deleted, however as can be seen in the center of Figure 7 all ten volume types are still present in the volume type data base.

Interestingly each of the ten volume types has a name of exactly 255 characters in length. The cinder type-create fuzz test created volume types with name lengths between 1 and 1025 characters. Volume types with name lengths between 1 and 254 as well as between 256 and 1025 characters were properly deleted using the OpenStack Horizon Dashboard.

International Journal on Cloud Computing: Services and Architecture (IJCCSA),Vol.2, No.6, December 2012In addition to using the OpenStack Horizon Dashboard to attempt to delete these ten volume types, the OpenStack cinder "type-delete" command was used to attempt to manually delete the ten image types. OpenStack cinder accepted, and appeared to execute each of the ten type-delete commands, however all ten image types were still in the database after the execution of the cinder type-delete commands.

There appears to be a problem in OpenStack Essex that prevents the deletion of a volume type with a name length of exactly 255 characters. To test this theory another volume type was created manually using the following cinder command:

> cinder --os_username admin
> --os_password adminpassword
> --os_tenant_name "admin"
> --os_auth_url http://192.168.1.10:5000/v2.0
> Type-create
>
> TestTestTestTestTestTestTestTestTestTestTestTestTestTestTestTestTestTestTestTestTestTestTestTestTestTestTestTestTestTestTestTestTestTestTestTestTestTestTestTeTestTestTestTestTestTestTestTestTestTestTestTestTestTestTestTestTestTestTestTestTestTes

When executed this command created a new volume type and assigned it a volume type number of 9527. The following cinder command was then used to delete the newly created volume:

> cinder --os_username admin
> --os_password adminpassword
> --os_tenant_name "admin"
> --os_auth_url http://192.168.1.10:5000/v2.0
> type-delete 9527

This command was executed with no errors. To verify that the volume type was deleted the following cinder command was executed:

> cinder --os_username admin
> --os_password adminpassword
> --os_tenant_name "admin"
> --os_auth_url http://192.168.1.10:5000/v2.0
> type-list

The new volume type did not show up in the list of volume types, so the cinder type-delete command appeared to work properly, even on a volume type with a name length of 255 characters. Based on these manual tests it appears that the problem is more complicated than not being able to delete a volume type with a name of exactly 255 characters.

The cinder file cinder/volume/volume_types.py" contains the python code to delete a volume type. [22] The portion of volume_types.py that implements the type-delete command is shown below.

```
def destroy(context, name):
    """Marks volume types as deleted."""
    if name is None:
        msg = _("name cannot be None")
```

56

International Journal on Cloud Computing: Services and Architecture (IJCCSA),Vol.2, No.6, December 2012

```
            raise exception.InvalidVolumeType(reason=msg)
        else:
            db.volume_type_destroy(context, name)
```

The python code for db.volume_type_destroy is located in the cinder/db/api.py file. [23] The portion of api.py that implements the type-delete command is shown below.

```
      def volume_type_destroy(context, name):
          """Delete a volume type."""
          return IMPL.volume_type_destroy(context, name)
```

A cursory analysis of the python code does not show any obvious bugs that could cause a volume type name of 255 characters in length to fail the delete process. The problem could be in the OpenStack MySQL database, or in other python code that is executed during the image-delete process. This problem has been submitted to the OpenStack Foundation as bug # 1085192. The bug submission has been accepted, assigned a high priority, and is planned to be resolved in the Grizzly-2 release early in 2013. The proposed solution is to use UUID based image type names rather than ASCII character based image type names.

### 7.2.2 Glance Command Line Fuzzing Test Results

Figure 8 shows a screen grab taken while the glance command line fuzzing test was running. The left side of the screen shows a Linux terminal executing the glance fuzz test shell script, while the right side of the screen shows the Ubuntu System Monitor program, which was used to monitor the CPU, Memory and Network utilization rates during the test. As shown in Figure 8, the average CPU utilization rate was less than 50% during this test. The OpenStack glance service did not crash, or exhibit any unexpected behaviours during the command line fuzz testing.

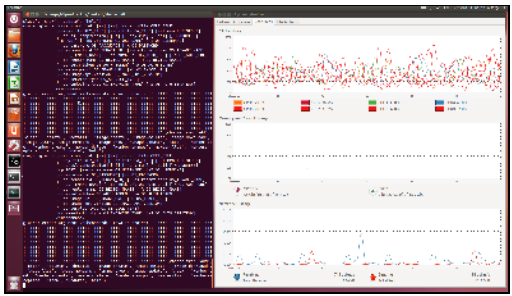
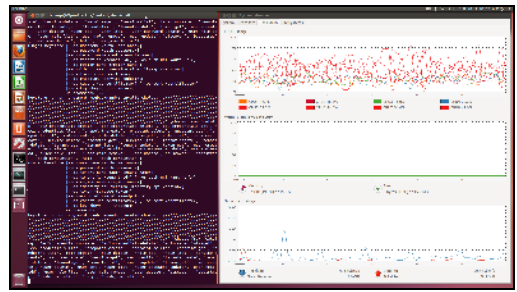

Figure 8: Glance Fuzz Test          Figure 9: Keystone Fuzz Test

### 7.2.3 Keystone Command Line Fuzzing Test Results

Figure 9 shows a screen grab taken while the keystone command line fuzzing test was running. The left side of the screen shows a Linux terminal executing the keystone fuzz test shell script, while the right side of the screen shows the Ubuntu System Monitor program, which was used to monitor the CPU, Memory and Network utilization rates during the test. As shown in Figure 9, the average CPU utilization rate was less than 50% during this test, although there were some bursts of utilization higher than 75%. The OpenStack keystone service did not crash, or exhibit any unexpected behaviours during the command line fuzz testing.

### 7.2.4 Nova Command Line Fuzzing Test Results

Figure 10 shows a screen grab taken while the nova command line fuzzing test was running. The left side of the screen shows a Linux terminal executing the nova fuzz test shell script, while the right side of the screen shows the Ubuntu System Monitor program, which was used to monitor the CPU, Memory and Network utilization rates during the test. As shown in Figure 10, the average CPU utilization rate was less than 50% during this test. The OpenStack nova service did not crash, or exhibit any unexpected behaviors during the command line fuzz testing.

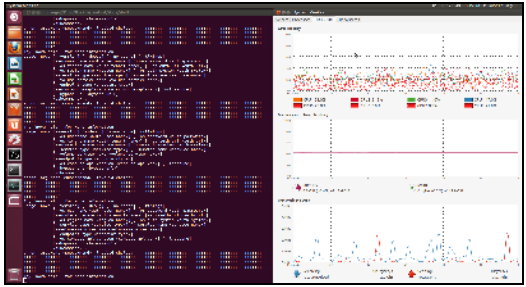 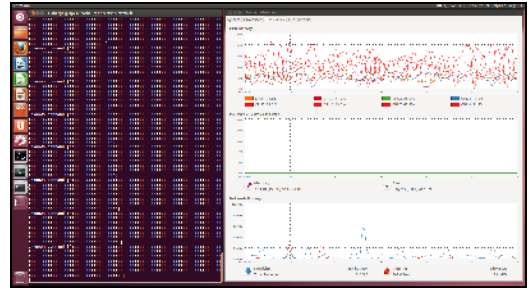

Figure 10: Nova Fuzz Test        Figure 11: Quantum Fuzz Test

### 7.2.5 Quantum Command Line Fuzzing Test Results

Figure 11 shows a screen grab taken while the quantum command line fuzzing test was running. The left side of the screen shows a Linux terminal executing the quantum fuzz test shell script, while the right side of the screen shows the Ubuntu System Monitor program that was used to monitor the CPU, Memory and Network utilization rates during the test. As shown in Figure 11, the average CPU utilization rate was less than 50% during this test, although there were bursts of CPU utilization above 75%. The OpenStack quantum service did not crash, or exhibit any unexpected behaviors during the command line fuzz testing.

### 7.2.6 Swift Command Line Fuzzing Test Results

Figure 12 shows a screen grab taken while the swift command line fuzzing test was running. The left side of the screen shows a Linux terminal executing the swift fuzz test shell script, while the right side of the screen shows the Ubuntu System Monitor program, which was used to monitor the CPU, Memory and Network utilization rates during the test. As shown in Figure 12, the average CPU utilization rate was less than 50% during this test. The OpenStack swift service did not crash, or exhibit any unexpected behaviors during the command line fuzz testing.

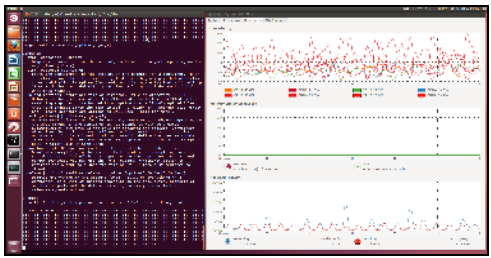 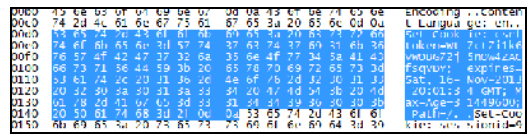

Figure 13: Session Cookie

Figure 12: Swift Fuzz Test

### 7.3. Session Hijacking

Figure 13 shows a portion of a Wireshark packet capture when OpenStack sends a session cookie to a Test_User_1 who is trying to access the Horizon Dashboard. As shown in Figure 13 the session cookie is transmitted as unencrypted ASCII text.

Figure 14 shows a screen shot of the Hamster program after it has read the session cookie information generated during Test_User_1's login session, and allowed an unauthorized user to access the Images & Snapshots web page of test-project-1.

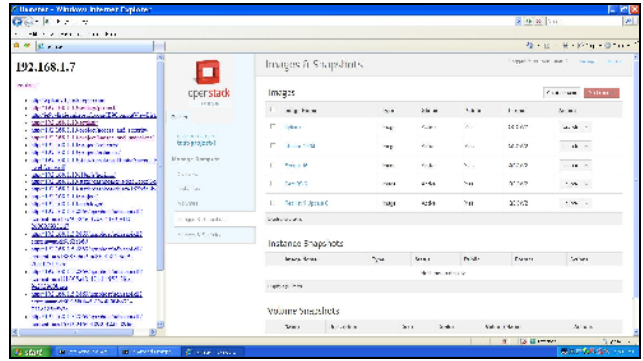

Figure 14: Test_User_1 Session Hijack

As discussed previously OpenStack released a patch to six Horizon python scripts to address a session hijacking vulnerability. The session hijacking penetration tests verified that these patches were installed on the OpenStack server, and that they properly prevented session hijacking after the user logged out.

During penetration tests these patches did not prevent session hijacking performed while an OpenStack user was still logged in. As long as the user was still logged in the Ferret and Hamster programs were able to hijack the user session. Once the session was hijacked an unauthorized user was able to access any of the restricted Horizon web pages associated with that user account. Access to these restricted web pages allowed an unauthorized user to add or delete cloud resources to any of the user's projects, to create snapshots of the user's instances or volumes, and to exfiltrate the user's instance or volume data.

This issue has been submitted to the OpenStack Foundation as bug # 1085198. The OpenStack Foundation is currently considering several options to address this issue. The most likely solution is to change the OpenStack documentation to make it clear that using a secure network protocol, like HTTPS, for communication between a cloud user and the Horizon Dashboard is strongly recommended. Horizon currently supports the HTTPS protocol, however in some cases it is not enabled during the OpenStack installation process.

### 7.4. Credential Theft

Figure 15 shows a portion of a Wireshark packet capture when a user logs into the Horizon Dashboard. As shown in Figure 19 the login credentials for Test_User_1 are transmitted as unencrypted ASCII text. In this case the user name, "Test_User_1", and the user's password, "password1", are transmitted over the network connection in the same unencrypted packet.

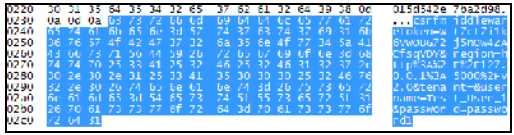

Figure 15: Test_User_1 Login

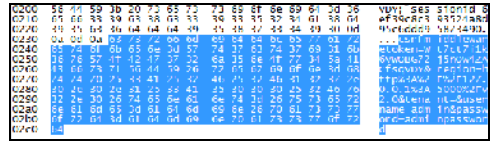

Figure 16: admin user login

Unlike the session hijacking vulnerability discussed above, this credential theft vulnerability can be exploited even after the user has logged out of their Horizon session. In fact this vulnerability can be exploited at any time of an attackers choosing provided the user has not changed their password. More importantly a patient attacker who has the ability to sniff the OpenStack server network connection could collect user login credentials for the OpenStack system administrator, and thus gain unauthorized access to all of the user data stored within the OpenStack server including images, instances, volumes, snapshots, and project data. Figure 16 shows a Wireshark capture of the OpenStack administrator's login credentials transmitted as unencrypted data over the network connection.

Analysis of the devstack, etc and var directories on the OpenStack server found several files where unencrypted administrative login credentials can be found. These files include:

- /devstack/localrc
- /etc/nova/api-paste.ini
- /etc/cinder/api-paste.ini
- /var/cache/cinder/cacert.pem
- /var/cache/cinder/signing_cert.pem
- /var/cache/glance/registry/cacert.pem
- /var/cache/glance/registry/signing_cert.pem

Since all of the files listed above are stored as unencrypted data it would be relatively easy for an insider threat to obtain access to these files and thus be able to steal OpenStack administrative credentials. It would be a bit more difficult for an outsider threat to access these files, but since OpenStack enables SSH access to the server a patient hacker could eventually crack the SSH password and then gain access to these files.

## 8. SUMMARY AND CONCLUSIONS

During this research effort a number of penetration tests were performed on an OpenStack Essex Cloud Management Server. HTTP Fuzzing of the OpenStack Horizon Dashboard user interface did not reveal any vulnerabilities or program errors. The HTTP fuzzing attacks used two freely distributed penetration test programs called BED and sfuzz.

Command line fuzzing of the OpenStack cinder service discovered a programming error related to deleting a volume type with a long file name (255 characters). Command line fuzzing of the OpenStack glance, keystone, nova, quantum and swift services did not reveal any vulnerabilities or programming errors. The OpenStack command line fuzzing attacks used the freely available sfuzz program.

A session hijacking attack against the OpenStack Horizon Dashboard service was successful and allowed an attacker to access restricted user information. The session hijacking attack used two freely distributed penetration testing programs called ferret and hamster. The session hijacking vulnerability is listed in the NIST National Vulnerability Database (CVE-2012-2144), and OpenStack has released a patch to address this vulnerability. Despite having the proper patches to address this vulnerability, a session hijacking attack is still possible under certain circumstances (i.e. the user whose session cookie was hijacked is still logged in to the OpenStack Horizon Dashboard).

Two different types of credential theft attacks were successful in allowing an attacker to learn a cloud user's or cloud administrator's login credentials, as well as to gain access to administrative certificates. Login credentials were acquired over an unencrypted network connection using the freely available Wireshark program. Administrative login credentials and certificates were acquired by locating unencrypted files on the OpenStack server that contained this sensitive information.

All of the vulnerabilities discovered during this research effort can be eliminated through the use of encryption. The session hijacking attack can be prevented by using HTTPS instead of HTTP for communications between cloud users and the cloud management software. The credential theft attacks can be prevented by encrypting any OpenStack files that contain sensitive information.

The programming error related to deleting volume types with long file names has been reported to the OpenStack foundation (bug # 1085192), been assigned a high priority, and is scheduled to be resolved in the Grizzly-2 release during early 2013.

The session hijacking issue has also been reported to the OpenStack Foundation (bug # 1085198). The most likely solution is to change the OpenStack documentation to strongly recommend the use of a secure protocol like HTTPS for network communications between cloud users and the Horizon Dashboard. OpenStack currently supports the HTTPS protocol, although it is not required for all installations.

It is important to continue to perform penetration tests on the OpenStack Cloud Management Software. OpenStack is being used by many large companies for their private, as well as public clouds. Improving the overall security posture of OpenStack through penetration testing is a worthy effort since many OpenStack users are moving more of their applications and data into the cloud